\documentclass[aps,pra,amsmath,amssymb,twocolumn]{revtex4-1}
\usepackage{fancyhdr}
\usepackage{graphicx}							
\usepackage{amssymb}
\usepackage{url}
\usepackage{steinmetz}
\usepackage[caption=false]{subfig}

\begin{document}
\title{Quantum optimal control via gradient ascent in function space and the time-bandwidth quantum speed limit }
\author{Dennis Lucarelli}
\email{dennis.lucarelli@jhuapl.edu}
 \affiliation{Johns Hopkins University Applied Physics Laboratory\\11100 Johns Hopkins Road, Laurel, MD, 20723, USA}

\begin{abstract}
A gradient ascent method for optimal quantum control synthesis is presented that employs a 
gradient derived with respect to  the coefficients of a functional basis expansion of the control.  
Restricting the space of allowable controls to weighted sums of the Slepian sequences efficiently parameterizes 
the control in terms of  bandwidth, resolution and pulse duration.  A bound showing minimum
time evolutions scaling with the inverse of the control bandwidth [S. Lloyd and S. Montangero, 
PRL, 113, 010502, (2014)] is recovered and the method is shown numerically to achieve the 
bound on entangling two-qubit quantum gates. 
\end{abstract}
\maketitle{}

\section{Introduction} 
Function families play a significant role in quantum mechanics often appearing as eigenfunctions
of the Schr\"{o}dinger equation.  Elementary examples include the Hermite polynomials that enter
as factors of the wavefunctions of the quantum harmonic oscillator, the Legendre and Laguerre polynomials 
associated with the spectrum of the hydrogen atom, and more recently the Mathieu functions as 
eigenfunctions to the Cooper pair box Hamiltonian.  

In the context of quantum information, functional analytic methods have been employed for 
quantum control, parameterizing the space of allowable controls and reducing 
the search space for optimal synthesis.  Recent results include the numerical determination 
of dynamically corrected gates using Walsh functions \cite{PhysRevA.84.062323, ball-biercuk},  the use of Hanning windows 
as a basis for frequency selective control of superconducting qubit devices \cite{PhysRevA.93.012324} and the control
protocol that uses randomized basis functions to parameterize the controls \cite{CRAB}.  These 
methods numerically locate optimal controls using a Nelder-Mead simplex method on the 
basis function coefficients. However, simplex based search methods generally scale poorly with the dimension of 
the unknown parameter \cite{nelder-mead-scaling}, thus limiting the number of basis functions in the expansion of the
control and the complexity of solutions amenable with this approach. 

Gradient based optimization methods, while converging only to a local maximum, scale well 
with problem size dimension and are commonly used in high dimensional optimization problems.  
The GRAPE algorithm \cite{Khaneja2005296} introduced a gradient based approach for solving the 
optimal control  problem for quantum gate synthesis.  Using the well-known formula for parameter 
differentiation of an exponential operator, GRAPE calculates a gradient at each time-step of the controlled evolution and
has been applied to a number of quantum control problems beyond its NMR origins. Calculating the gradient at
each time-step independently, however, can lead to widely varying control amplitudes at sequential times potentially 
violating limits on the slew rate or {\it bandwidth} of the control.  

This paper introduces a method, 
{\it Gradient Ascent in Function Space} (GRAFS), that leverages standard gradient ascent solvers on the set of basis function 
coefficients parameterizing the entire time extent of the control.  This functional analytic approach incurs little additional
computational effort over GRAPE and offers a number of advantages, most notably, a significant dimensionality reduction of the underlying
optimization problem and the ability to constrain the control manifold to smooth, well-behaved function families 
chosen to match the control problem at hand. The GRAFS method generalizes a gradient expression derived for 
control of spins in NMR \cite{Skinner2010248} and subsequently used for optimal control of classical systems \cite{Meister2014}.  
The gradient  derived here is seen as a consequence of the product rule applied to the 
matrix product defined by a first-order Trotter expansion of the propagator, thus generalizing the 
functional gradient in \cite{Skinner2010248} beyond two-level semi-classical systems. 

Similarly, a recent method \cite{GOAT} constructs high-fidelity, {\it analytic} controls expressed in 
a basis function expansion. In that work, a dynamical equation is derived for the gradient of the
propagator with respect to its parameters that couples to the Schr\"{o}dinger dynamics. The adjoined 
system is then integrated to construct the gradient and used to maximize the objective function. 

This paper also investigates the role of control bandwidth 
in quantum control problems.  Constraints on the control bandwidth arise from practical considerations of
experimental procedures and classical control electronics and are present in all modern qubit  systems.
With the focus on bandwidth, it is natural to consider the  \textit{discrete prolate spheroidal sequences} \cite{pwsf_5},
commonly referred to as the {\it Slepian sequences}, as a basis for piece-wise constant controls. 
These finite length sequences faithfully and efficiently represent the space of 
band-limited signals and serve as the \textit{basis functions} parameterizing the controls in this paper. 
Given the experimental constraints imposed on quantum controls, methods have been proposed to
account for the effects of band-limited, bounded control in optimal control synthesis 
\cite{PhysRevA.84.022307, Pawela2013, PhysRevApplied.4.024012,  Heeres2017}.  These methods introduce
a  penalty on the derivative of the control term in the control objective function, a Fourier cut-off constraint or 
a band-limiting filter in the control protocol iterates.  The gradient method proposed here, with the 
aid of the Slepian sequences, constrains the space of controls {\it a priori} to produce an intrinsic 
control solution without additional terms in the objective function or {\it post hoc} band-limiting. 

Closed quantum control systems have a convenient representation as control systems on Lie groups 
\cite{brockett_72, JurdjevicSussman} from which a number of properties can be immediately inferred. In particular, 
the \textit{Lie algebra rank condition} (LARC), determined by the Lie algebra spanned by repeated commutators 
of the internal (or drift) Hamiltonian and the control Hamiltonians,  can be used to determine the \textit{reachability} 
of a quantum system. The reachability condition ensures the existence of controls that can reach any point 
in state space  and is a generic condition for systems evolving on (compact) Lie groups \cite{BOOTHBY1975296, PhysRevLett.75.346}.
The LARC, however, does not lead to a constructive procedure for determining the controls required to 
drive the system to a desired element of the Lie group.  This {\it control synthesis} problem requires a solution to a two-point
boundary value problem and can solved analytically by pulse area theorems or in special cases (c.f. \cite{PhysRevB.91.161405}).  
For more general, complex control situations, numerical methods from optimal control theory are often necessary. 

Fundamental  limits on the evolution time of a quantum system are often stated as {\it quantum speed limits} (QSL). 
These bounds are often geometric restatements of the time-energy uncertainty relation for time independent systems 
\cite{ M-T-bounds, PhysRev.122.1649, MARGOLUS1998188} and have been generalized to time-dependent, driven 
quantum systems \cite {PhysRevLett.111.010402, PhysRevLett.110.050402,PhysRevLett.110.050403}.  Due to relatively
short coherence times and slow logic gates, the QSL is an important limit in multiple qubit quantum gate synthesis \cite{goerz_koch}. 
Recently, the set of quantum states reachable in polynomial time with bounded controls has been 
characterized and an associated bound has been derived relating the control 
bandwidth, final state accuracy and minimum achievable evolution time \cite{lloyd_qsl}.  
This {\it time-bandwidth} QSL has an intuitive appeal -- higher bandwidth controls drive a quantum system
to its target state at a faster rate.  The gradient method introduced here approximately realizes this bound
and thus demonstrates a control synthesis method that elucidates the limits of bounded, band-limited control.

The band-limited characteristics of Slepian-modulated controls as applied to qubits have recently been 
experimentally validated and deployed for narrowband spectroscopy of noise in a microwave control
system \cite{Frey2017}. Further results and details on Slepian based noise spectroscopy can be found
in the recent manuscript \cite{1803.05538}.

\section{The space of time-band-limited sequences}
The time-bandwidth uncertainty relation governs the concentration properties of
classical signals.  In the limiting case, the relation is observed by a pure tone with infinite time extent and 
vanishing spectral measure. As a purely mathematical object, the only signal that is both time and bandwidth
limited is trivially zero. Motivated by problems in communications technology, where all 
signals are manifestly both time and band-limited, a theory of band-limited functions was developed by 
seeking a engineering compromise to the uncertainty relation
 \cite{pwsf_1, pwsf_2, pwsf_3, pwsf_4}. These functions are characterized by maximizing the fraction
 of a signals energy in a time interval while enforcing a band-limiting constraint.  The solution to this optimization
 problem was aided by the discovery of a differential operator that simultaneously commutes with time and band-limiting
 operators. The form of this commuting operator is related to the wave equation in prolate spheroidal coordinates, thus
 christening the resulting functions as the {\it prolate spheroidal wave functions} (PSWF).  
 
From a signals and control perspective, it is often useful to consider the discrete analogues to the PSWF, the 
 {\em discrete prolate spheroidal sequences} or simply the Slepian sequences \cite{pwsf_5}. 
For a sequence of length $N$ and half bandwidth $W,$ the Slepian sequences, $v_k( N,W),$ 
are defined as real solutions to the eigenvalue problem
\begin{equation}
\sum_{m=0}^{N-1} \frac{\sin 2\pi W(n-m)}{\pi(n-m)} v_k(m; N,W) = \lambda_k(N,W) v_k(\ell; N,W)
\end{equation}
where $k,\ell \in \{0, 1, \dots, N-1\}$ and $v_k(\ell; N,W)$ is the $\ell$-th element of $k$-th order Slepian.  
The eigenvalues $\lambda_k(N,W)$ can be shown to be a measure of 
 spectral concentration and are ordered such that 
 $1 \geq \lambda_0(N,W) \geq \lambda_1(N,W) \geq \dots \geq \lambda_N(N,W)$
and have the remarkable property that the first $2NW$ eigenvalues are near unity, 
while the remaining eigenvalues are close to $0.$ In Ref. \cite{slepian_comments}, Slepian uses this 
property to  establish the approximate dimension of the space of band-limited sequences of 
length $N$ to be $2NW.$ The  Slepian order, $k,$ for the sequence $v_k( N,W),$ indicates the
even/odd symmetry of the sequence with respect to its midpoint and its number of zero-crossings.
In the following, weighted sums of Slepian sequences are identified as piecewise constant controls by specifying a 
sequence length $N,$ a half bandwidth parameter $W \in (0,0.5)$ and a pulse duration $\tau.$ 
The pair $\left(N,\tau\right)$ then defines a control resolution $\Delta t,$ the associated 
Nyquist frequency and $W/\Delta t$ is in units of hertz.

\section{\label{sec:GRAFS}Quantum Optimal control and Gradient Ascent in Function Space}
For a closed, finite dimensional quantum system,  the state dependence of the Schr\"{o}dinger equation is often removed and
the dynamics are equivalently lifted to a differential equation on the space of special unitary matrices $U \in SU(d), $ 
that act on the $d-$dimensional Hilbert space of quantum states.  In this  representation, the quantum control system is given by
\begin{equation}\label{system1}
\dot{U}(t) = -\frac{i}{\hbar} \left(  H_d + \sum_{j=1}^M \Omega_j(t)H_j \right) U(t)
\end{equation} 
where $H_d$ is the time-independent drift Hamiltonian and the $H_j$ are the control Hamiltonians
that capture the coupling of the qubits to an externally applied control field.  Denote the controlled Hamiltonian with drift as 
\begin{equation}\label{H_Omega}
H_\Omega(t) = H_d + \sum_{j=1}^M \Omega_j(t)H_j 
\end{equation}
The solution to (\ref{system1}) must account for the non-commuting operators 
$\left[ H_\Omega(t_\ell) \, , \, H_\Omega(t_\ell) \right]  \neq 0 \, \, \text{for} \, \, k \neq \ell $
and is formally given by the time-ordered integral
\begin{equation}\label{time-integral}
U_\tau(\Omega) =  {\mathbf T} \exp \left[ -\frac{i}{\hbar} \int_0^\tau  H_\Omega(t) dt\right]
\end{equation}
With initial conditions given by the identity matrix,  $U_0 = \boldsymbol{1} \, , $ 
the solution of (\ref{system1}) simply propagates the state by matrix-vector 
multiplication $\psi_\tau = U_\tau \cdot \psi_0 \, . $ 

\begin{figure}[t!]
\includegraphics[width = 3.35in]{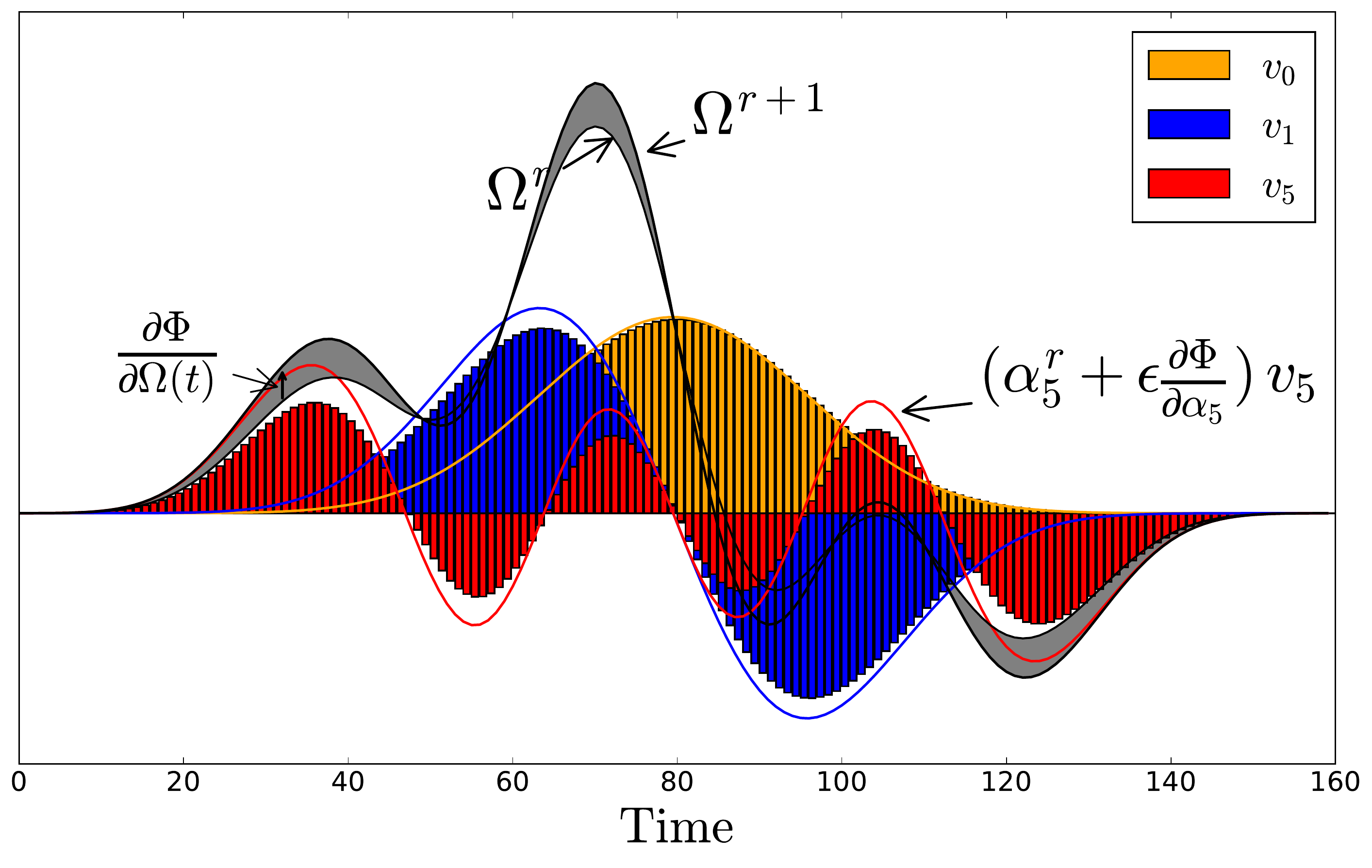}
\caption{(Color online) Depiction of GRAPE and GRAFS control iterates.  GRAPE updates the control 
at each time-step ($\frac{\partial \Phi}{\partial \Omega(t)}$) , while GRAFS increments the basis function 
coefficient and affects the control at all times ($\frac{\partial \Phi}{\partial \alpha}$). 
Slepian sequences,  $v_k,$ are shown for $k=0,1,$ and $5$}. 
\label{grape_and_GRAFS}
\end{figure}

The quantum gate synthesis problem is to determine the control functions $\left\{ \Omega_j(t) \right\}_{j=1}^M$ 
that generate a desired target unitary $U_{targ}$ at some time $\tau \, . $ Optimal control typically defines an 
objective function that is optimized subject to a dynamical constraint and initial and final conditions on the 
state of the system. In quantum gate synthesis, this methodology is employed to determine control fields 
that minimize trace distance from a target unitary while evolving  according to the Schr\"{o}dinger equation (\ref{system1}).  

The optimal control objective function may take several forms depending on the control task, but all 
require a fidelity measure, ${\cal F}$, to be maximized.  A common choice is the (global) phase invariant distance to 
the target unitary at the at the final time $\tau$

\begin{equation}\label{phase-invariant}
\Phi({\cal F} (\Omega))=  \frac{1}{d}\left | {\cal F}(\Omega) \right |
\end{equation}
normalized by the dimension of the system, $d$, with the trace fidelity measure
\begin{equation}
{\cal F}(\Omega)  = \mathrm{tr} \left[ U_{targ}^\dagger \cdot U_\tau(\Omega) \right]
\end{equation}
First order optimality conditions, known as  
\textit{Pontrayagin's maximum principle} (PMP), are well established but, in general, result in a system of 
nonlinear  differential equations that must be solved for optimal control synthesis.   
In  Ref. \cite{Khaneja2005296}, a \textit{Gradient Ascent Pulse  Engineering} algorithm (GRAPE) was 
introduced to numerically solve the optimal control problem that simplifies the optimization problem 
by iteratively updating the control fields by small steps in the gradient direction of the objective 
function at each time-step of the control.  The pulse shaping iterates are of the form
\begin{equation}\label{grape_iterates}
\Omega_j^{r+1}(t_\ell)  = \Omega_j^{r}(t_\ell) + \epsilon \, \frac{\partial \Phi(\Omega^{r})}{\partial \Omega_j(t_\ell)}
\end{equation}
with $r$ denoting the iteration number and the increment $\epsilon$ being chosen by line search.  
While consistent with the PMP, GRAPE derives its
algorithmic simplicity by removing any explicit constraints on the control in the objective 
function.  For a system without integral constraints on the controls, 
the PMP conditions result in trivial {\em co-state} dynamics, simplifying the first order
necessary conditions  \cite{pontryagin_book} and ensuring that gradient ascent updates 
are sufficient for convergence to a local maximum. 

\begin{figure*}[t!]
\subfloat{
\includegraphics[height=2.2in]{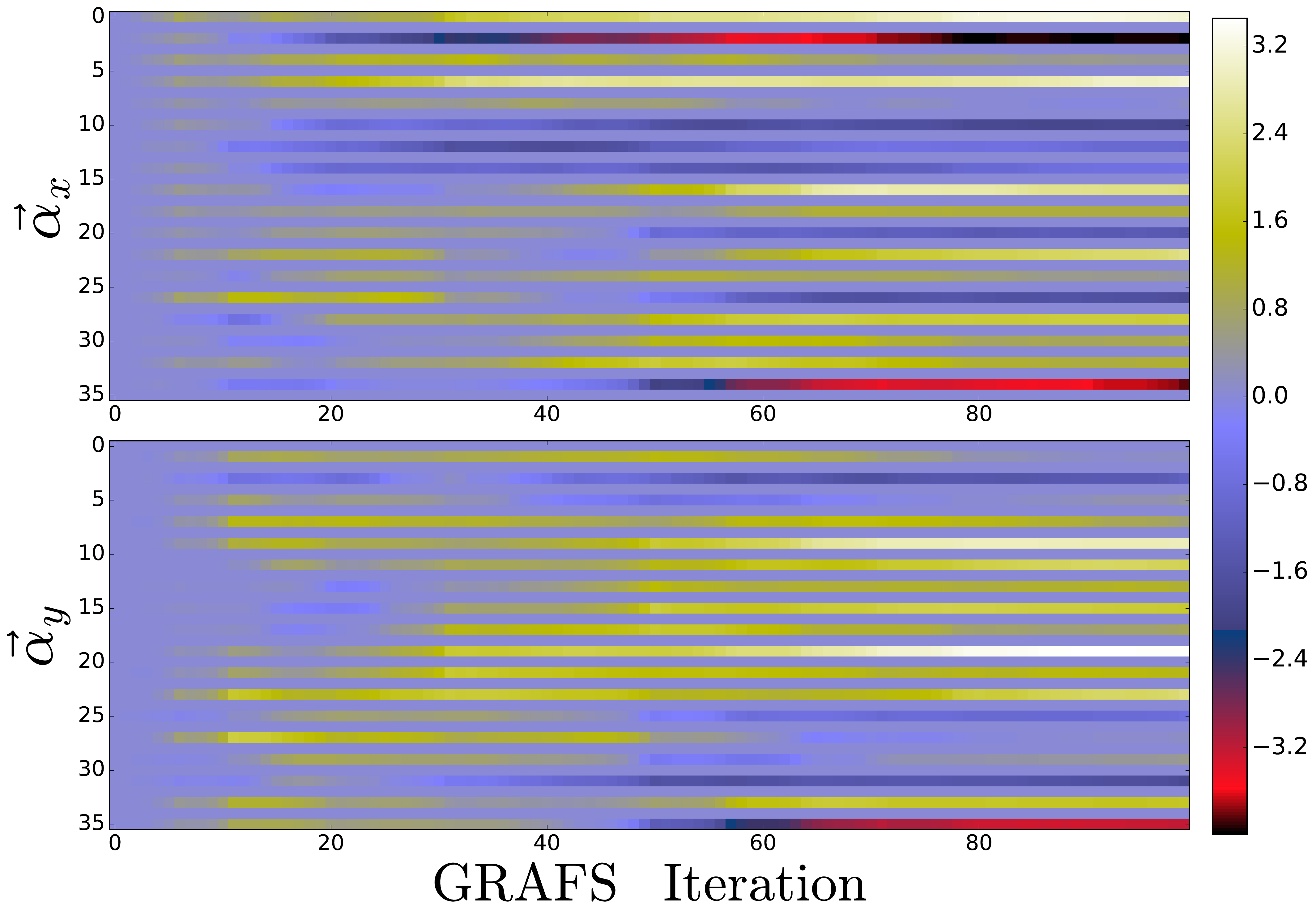}
} \, \, 
\subfloat{
\includegraphics[height = 2.2in]{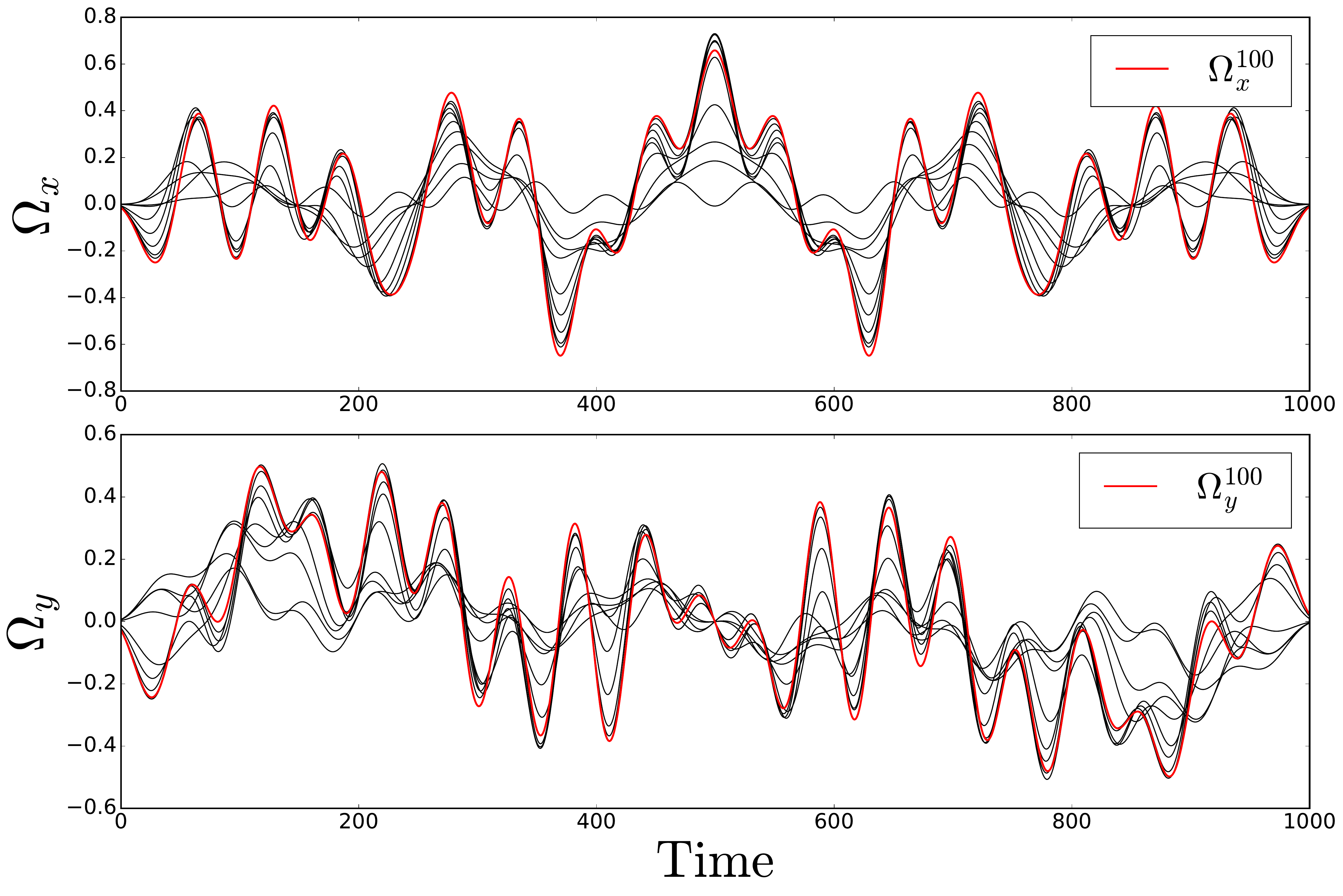}
}
\caption{(Color online) GRAFS iterations and convergence for constructing a Toffoli gate 
with two single qubit controls and Heisenberg exchange operations.  The optimization trajectory of Slepian sequence weights is 
shown in the left panel and 10 pulse shaping iterations are shown on the right where  
$\Omega^r_j = {\cal V} \cdot {\vec{\alpha_j}}^r \, . $ The final pulse shape is shown in bold red.}
\label{toffoli-iterations}
\end{figure*}

Assuming a discretization of the controls into $N$ equal $\Delta t$ length segments with total
pulse duration, $\tau = N \Delta t,$ the GRAPE 
algorithm proceeds by using the \textit{product of exponentials} 
\begin{equation}
\label{product-of-exponentials}
U_\tau \approx e^{{-\frac{i \Delta t}{\hbar}}H_\Omega(t_N)} e^{-\frac{i \Delta t}{\hbar} 
H_\Omega(t_{N-1})}\cdot \cdot \cdot  e^{-\frac{i \Delta t}{\hbar} H_\Omega(t_{1})}
\end{equation}
as a good approximation to the time-ordered integral (\ref{time-integral}) and the exact 
diagonalization formula \cite{aizu} for parameter differentiation of the matrix
exponential  
\begin{equation}
\label{parameter-diff}
\frac{d}{ds}  e^{A + s B} 
\end{equation} 
Let $\{ |\lambda_\nu\rangle \, , \lambda_\nu \}$ denote an eigensystem for the matrix $A + sB$, then the
matrix elements of the partial derivative are given by
\begin{align}\label{exp-partial}
\langle \lambda_\nu | &\frac{\partial [A + sB]}{\partial s} | \lambda_\mu \rangle = \\
&\begin{cases} \nonumber
-i\Delta t \langle \lambda_\nu | B |   \lambda_\nu \rangle \, e^{-i\Delta t \lambda_\nu} \quad &\mathrm{for} \, \, \nu=\mu \\ 
-i\Delta t \langle \lambda_\nu | B |   \lambda_\mu \rangle 
\, \frac{e^{-i\Delta t \lambda_\nu} - e^{-i\Delta t \lambda_\mu} }{\lambda_\nu - \lambda_\mu}  &\mathrm{for} \, \, \nu\neq\mu
\end{cases}
\end{align}
Define the notation 
\begin{equation}
U_{k:\ell}  = U(t_{k}) \cdots U(t_\ell) 
\end{equation}
for matrix products of the form (\ref{product-of-exponentials}). Taking the partial derivative of  
(\ref{product-of-exponentials}) with respect to the $j$-th control at time $t_\ell = \ell\,\Delta t $ yields
\begin{equation}\label{partial-time}
\frac{\partial U_\tau}{\partial \Omega_j(t_\ell)} = U_{N:\ell+1}\,\frac{\partial U(t_\ell)}{\partial \Omega_j(t_\ell)}\, U_{\ell-1:1}
\end{equation}
and by the chain rule we have
\begin{equation}\label{partial_phi_time}
\frac{\partial \Phi(\Omega)}{\partial\Omega(t_\ell)}  = \frac{1}{d}\,\mathrm{Re} \left[e^{-i \arg({\cal F})}  \mathrm{tr}[U_{targ}^\dagger \cdot \frac{\partial U_\tau}{\partial \Omega_j(t_\ell)} ] \right] 
\end{equation} 
where $\arg(\cdot)$ denotes the phase of a complex number.  These expressions and 
their derivations can be found in \cite{PhysRevA.84.022305}. Denote the GRAPE gradient 
over all controls as the multi-dimensional tensor 
$ \nabla_{\Omega} U_\tau \in  \mathbb{C}^{N\times M \times d \times d} $ with 
components $ \nabla_{\Omega_j}^\ell U_\tau $  given by (\ref{partial-time}).

In contrast to updating the controls at each time-step, the GRAFS method first expresses  
the controls in a functional basis expansion and exploits the observation that the gradient of 
the propagator at the final time with respect to the  the basis function coefficients is simply 
an application of the product rule on the matrix products defining the approximate propagator. 
The update rules are depicted in \figurename{~\ref{grape_and_GRAFS}}, 
with the essential difference being that variations of the basis function coefficients globally 
affect the control field over all times, whereas GRAPE updates affect each time-step independently. 

The controls are formally expressed as weighted sums of length $N$ piecewise constant sequences $v_k(t)$
\begin{equation}
\Omega_j(t_\ell) = \sum_{k=0}^K\alpha_{kj} v_k(t_\ell)
\end{equation}
with real coefficients $\alpha_{kj}$.  While the method is applicable to any set of basis functions, 
the simulations in the following are performed with controls expressed as weighted sums 
of Slepian sequences.  Applying the product rule to the matrix product (\ref{product-of-exponentials}) 
and setting $\hbar =1$,  the partial derivative is given by
\begin{equation}\label{GRAFS_partial}
\frac{\partial{U_\tau}}{\partial \alpha_{kj}} = \sum_{\ell=1}^N U_{N:\ell+1} \frac{\partial U(t_\ell)}{\partial \alpha_{kj}} U_{\ell-1:1}
\end{equation}
where 
\begin{equation}\label{alpha-partials}
\frac{\partial U(t_\ell)}{\partial \alpha_{kj}} = \frac{\partial \exp[ -i \Delta t H_\Omega(t_\ell) -i \alpha_{kj} \Delta t v_k(t_\ell) H_j]}{\partial \alpha_{kj}} 
\end{equation} 
is of the form (\ref{parameter-diff}) with 
$A = -i \Delta t H_\Omega(t_\ell)\, , \, B = -i \Delta t v_k(t_\ell) H_j $ and $s = \alpha_{kj}.$
Note that by the formula (\ref{exp-partial}), the scalars $v_k(t_\ell)$ distribute out of the inner product to yield
\begin{equation}\label{dist-v}
\frac{\partial U(t_\ell)}{\partial \alpha_{jk}} = v_k(t_\ell) \,  \frac{\partial U(t_\ell)}{\partial \Omega_j(t_\ell)}
\end{equation}
Let ${\cal A}$ denote the matrix of coefficients so that  ${\cal A}_{kj} = \alpha_{kj}$ and
arrange the basis functions as columns of the $N \times K$  matrix ${\cal V}\, .$ Using (\ref{dist-v}), the GRAFS gradient,
$\nabla_{\cal A} U_\tau  \in \mathbb{C}^{K\times M \times d \times d} \, ,$ 
can be viewed as a contraction over the time index
\begin{equation}\label{tensor-dot-grad}
\nabla_{\cal A} U_\tau = {\cal V} \otimes_\ell \nabla_\Omega U_\tau 
\end{equation}
with the matrix-valued components 
\begin{equation}
\left[\nabla_{\cal A} U_\tau\right]_{kj} = \sum_{\ell=1}^N {\cal V}_{\ell k} \,\nabla_{\Omega_j}^\ell U_\tau 
\end{equation}
This {\em tensor-dot} operation can be efficiently implemented with the matrix 
of basis functions and the GRAPE gradient and thus the GRAFS gradient 
incurs little additional computational overhead.  Finally, the chain rule (\ref{partial_phi_time}) 
can be applied to obtain $\nabla_{\cal A} \Phi$ and the gradient ascent control update in matrix form is simply
\begin{equation} \label{matrix_update}
\Omega^{r+1} = {\cal V} \cdot \left({\cal A}^{r} + \epsilon \,\nabla_{{\cal A}^r} \Phi\right)
\end{equation}

An advantage of GRAFS is that control constraints 
are enforced \textit{a priori}, through the use of basis functions, while still employing the gradient ascent procedure. 
Since arbitrary pulse shapes can often be faithfully represented by a small set of basis functions, GRAFS also constructs 
a more \textit {efficient} optimization problem.  In the case of Slepian parameterized controls, $2NW$ sequences
approximately characterize the space of band-limited sequences.  For a system with $M$ controls, this results
in an unknown parameter vector of dimension $2NWM$ for GRAFS compared to $NM$ for GRAPE.  
Since $W$ is strictly (and typically much) less than $0.5$, this represents a significant reduction in dimensionality of the 
underlying optimization problem.  

An illustrative example of the method is given by the following notional quantum control system.  
Suppose the control task is to construct a three qubit Toffoli (or controlled-controlled-not) gate from 
Heisenberg exchange operators and two independent controls. Denoting the Heisenberg exchange on 
qubits $i$ and $j$ as $H_{\cal X}^{ij}$ and a single qubit control on qubit $k$ as
$\sigma^k$ the system Hamiltonian is given by 
\begin{equation}\label{3-qubit}
H_\Omega = H_{\cal X}^{12}  + H_{\cal X}^{23}  + \Omega_x(t) \sigma_x^1 
+ \Omega_y(t)\sigma_y^3
\end{equation}

\begin{figure}[t]
\includegraphics[width=3.5in]{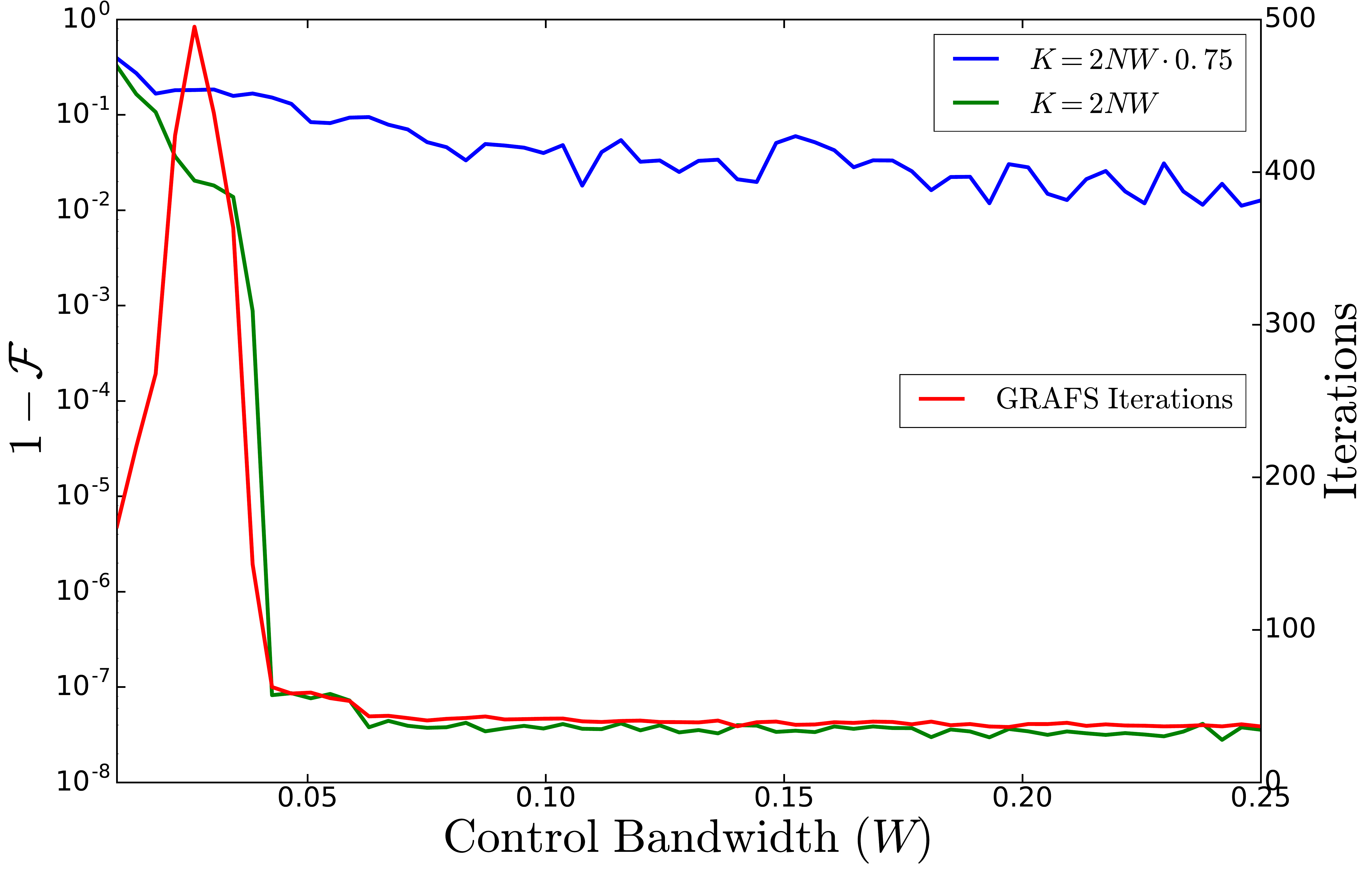}
\caption{(Color online) Infidelity as a function of bandwidth (W)  and number of Slepians sequences (K).  
Also shown is the number of GRAFS iterations to convergence}
\label{slep_scaling}
\end{figure}

\figurename{ \ref{toffoli-iterations}} shows the optimization trajectory of 100 iterations of the 
GRAFS algorithm using the L-BFGS algorithm to impose inequality constraints
on the basis function weights $(|\alpha_{kj}| < 5.0)$ and perform line search to calculate 
$\epsilon$ in the gradient update (\ref{matrix_update}).  Controls were formed with a Slepian basis expansion 
with parameters $N = 1000$ and $W=0.02. $  Initial conditions
for the coefficients for were set to zero. With these parameters, there are 40 Slepian sequences. 
Higher-order Slepians hold less spectral concentration and can have non-zero initial and final points. 
Removing these sequences leaves a final set of 36 sequences for pulse shaping. 
The final infidelity after 100 iterations was ${\cal O}(10^{-7})$. 
All simulations were performed with the optimal control module 
in Qutip \cite{qutip}  by modifying the gradient calculation to 
(\ref{tensor-dot-grad}) and the update equation to (\ref{matrix_update}). 

\begin{figure}[t]
\includegraphics[width=3.5in]{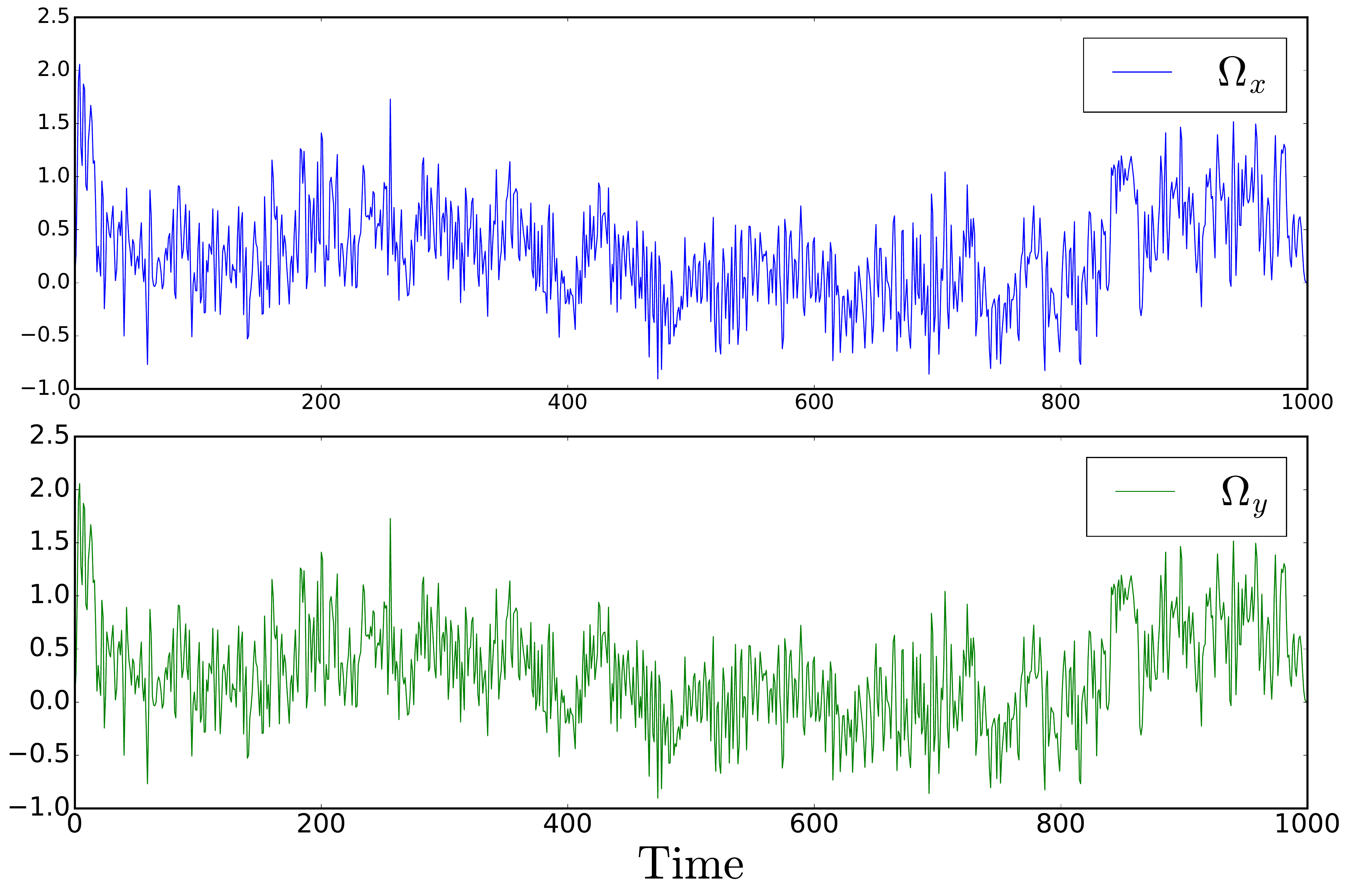}
\caption{(Color online) High bandwidth controls $(W=0.4)$ for system (\ref{3-qubit}) }
\label{high_bw}
\end{figure}

Intuitively, controls constructed from high bandwidth Slepian sequences should contain more control 
authority and thus result in synthesis of high fidelity gates. \figurename{ \ref{slep_scaling}} 
confirms this intuition and also shows the importance of using a nearly complete set of $K=2NW$ 
Slepian sequences in the basis function expansion of the control.  By varying the bandwidth 
and running the GRAFS procedure with the system (\ref{3-qubit}), high fidelity Toffoli gates could not be 
generated across the entire bandwidth range ($W \in [0.01,.25]$ shown)when using just three quarters of the available sequences. 
As shown in \figurename{ \ref{slep_scaling}}, convergence of the algorithm, as determined by the norm of the gradient getting small $(10^{-9})$, also
scales with the bandwidth of the control. High bandwidth controls, constructed from a Slepian basis 
with $N=1000, \, W=0.4$ and the GRAFS method,  are shown in \figurename{ \ref{high_bw}}.  
These physically unrealizable controls approach the {\it bang-bang} limit and demonstrate how
the Slepian sequences, with their continuous bandwidth parameter, represent the full space of controls.

\section{Time-bandwidth Quantum Speed Limit}
Further analysis reveals a time-bandwidth relation by considering the QSL
for time-dependent open quantum systems 
\cite{PhysRevLett.110.050403, PhysRevLett.111.010402, PhysRevLett.110.050402}
\begin{equation}\label{QSL}
{\cal T}_{QSL} \geq \frac{\delta}{\frac{1}{\tau}\int_0^\tau ||H_\Omega(t)||_p \,dt}
\end{equation}
where $\delta $ is an accuracy measure and $|| \cdot ||_p$ is the matrix $p$-norm.  
For mixed state transfer from $\rho_0$ to $\rho_f$, the accuracy measure is given in terms of the Bures angle 
$\delta = \sin^2\left( \arccos \sqrt{\mathrm{tr} \left [ \sqrt{\rho_0} \rho_f \sqrt{\rho_0} \right]} \right) \, .$ 

Restricting the space of allowable controls to the Slepian sequences and assuming closed, unitary dynamics
and setting $p=2,$  a bound on the denominator of (\ref{QSL}) can be derived. Considering a single control Hamiltonian $H_c,$ 
no drift term and denoting  $ {\overline H_\Omega} = \frac{1}{\tau} \int_0^\tau ||H_\Omega(t)||_2  $  and $\alpha = \max_i \{\alpha_i\}$
the following string of inequalities can be obtained
\begin{eqnarray}
{\overline H_\Omega} &\leq& \frac{1}{\tau}\sum\limits_{i=1}^{2NW}  \int_0^\tau  || \alpha_i v_i(t) H_c ||_2 \, dt\\
&=& \frac{1}{\tau}||\alpha \, H_c ||_2 \sum\limits_{i=1}^{2NW}  \sum_{\ell=1}^N |v_i(t_\ell)|\Delta t  \\
&\leq& \frac{1}{\tau}||\alpha \, H_c ||_2 \sum\limits_{i=1}^{2NW}  \sqrt{N}\Delta t || v_i ||_2 \\ 
&=& \frac{1}{\tau}||\alpha \, H_c ||_2  \sqrt{N}\Delta t 2NW
\end{eqnarray}
since  $||v_i||_1 < \sqrt{N} ||v_i||_2$ and the Slepian sequences are normalized in two-norm.
Setting $|\alpha| = ||H_c||_2 = 1$ and since $\tau = N\Delta t$, a time-bandwidth QSL similar to Eq. 11 in Ref. \cite{lloyd_qsl} 
\begin{equation}\label{BW-QSL}
{\cal T}_{QSL} (2\sqrt{N}W) \geq \delta
\end{equation}
is easily recovered.

\begin{figure}[t!]
\includegraphics[width=3.5in]{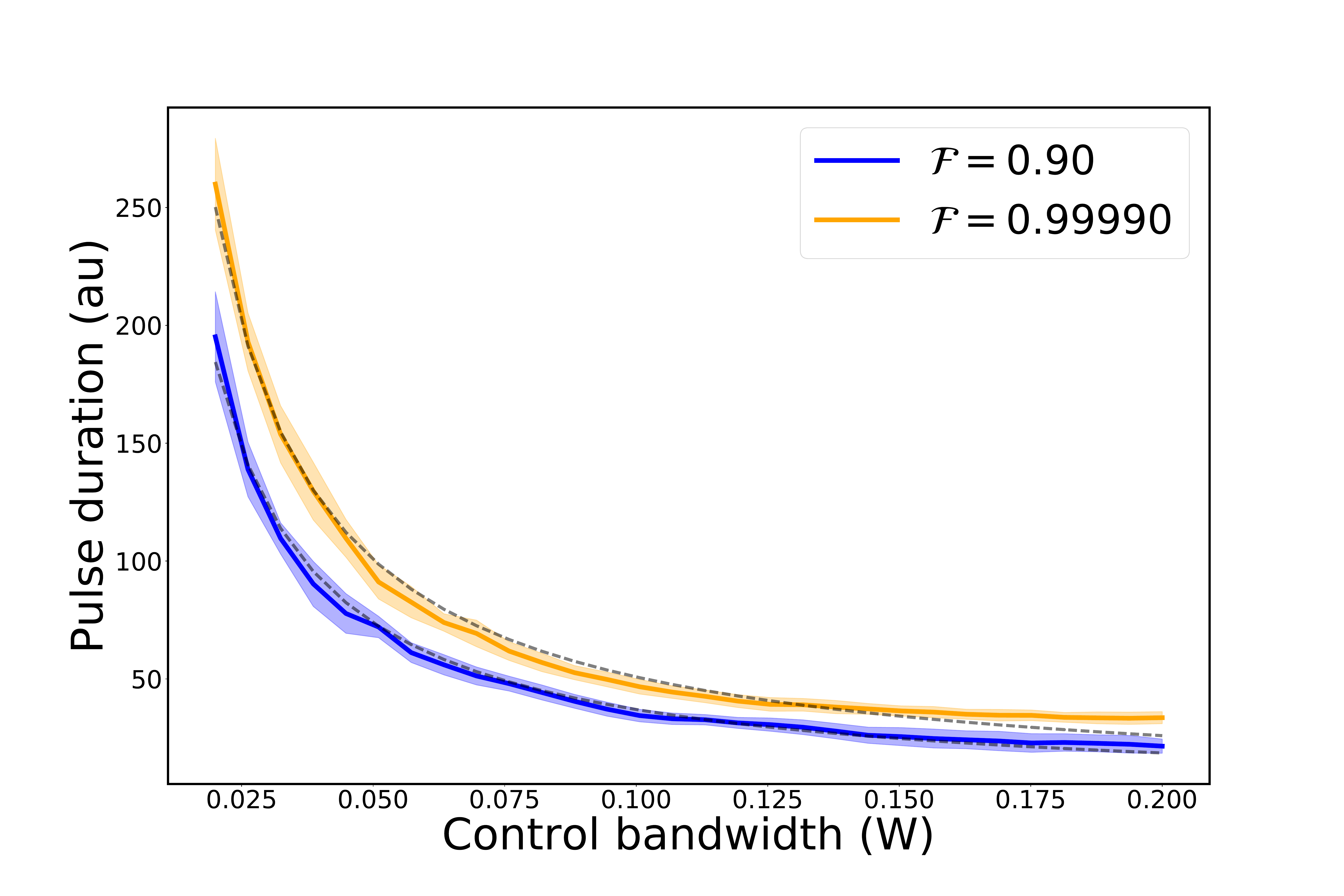}
\caption{(Color online) Empirical Quantum Speed Limit scaling for ${\cal F} = 0.90$ (blue) 
and ${\cal F}=0.99990$ (orange). The fill regions denote the standard deviation across 
all 100 samples. Dashed lines are curve fits of the form $\frac{a}{W} + b$}
\label{QSL-scaling}
\end{figure} 

A series of numerical experiments were performed to qualitatively assess the 
${\cal T}_{QSL}  \geq {\cal O}(\delta/W) $ scaling by running the GRAFS procedure 
repeatedly to determine the minimal time to achieve a specified fidelity given by the distance measure
$\delta=\Phi({\cal F})$. In these experiments, 100 target unitaries were constructed by random 
sampling parameters defining two-qubit gates in the Cartan decomposition. In particular, a 
tuple $(c_{x}, c_{y}, c_{z})$ is sampled  from the Weyl chamber \cite{PhysRevA.67.042313}  to form the target 
\begin{equation}
U_{targ} = K_1 \exp\left(\frac{i}{2} \left( c_{x}\sigma_x \sigma_x + c_{y}\sigma_y\sigma_y +c_{z}\sigma_z \sigma_z  \right)\right) K_2
\end{equation} 
with randomly generated local operations $K_i \in SU(2)\otimes SU(2)\, .$ To further constrain the set of target gates, 
the tuples are formed by sampling from the Weyl chamber to obtain parameters from the 7-faced
polyhedron defining parameters for so-called \textit{perfect entanglers} \cite{PhysRevA.67.042313}. These are the two-qubit unitaries 
capable of  generating maximally entangled states from an initial product state.  This class of unitaries contains the well known
quantum logic gates such as the  CNOT and $\sqrt{\mathrm{SWAP}} \, .$

GRAFS was used to determine controls for this set of target gates while varying the control bandwidth, $W,$
and keeping all other parameters, $\max_i \{|\alpha_i|\} = 1$ and $N=1000,$ fixed.
The two-qubit Hamiltonian system with 4 independent controls is given by
\begin{equation}\label{2q}
H^{12}_{\cal X} + \sum_{i=1}^2 \Omega^i_x(t) \sigma_x^i + \Omega^i_y(t)\sigma_y^i
\end{equation}

A bracketing procedure, requiring multiple runs of the algorithm for each target gate, was 
used to estimate a minimal pulse duration for a given set of parameters and desired fidelity.  
Specifically, given a perfect entangler target, nominal control bandwidth, $W,$ and pulse
duration, $\tau$, the GRAFS algorithm is run until convergence or until reaching the desired fidelity. 
The algorithm maintains and updates three candidate times with differences forming a golden ratio. If 
the desired fidelity if obtained, the algorithm is run again with the same target but with a pulse
duration decreased by the distance scaled by the golden ratio. If the algorithm converges without 
reaching the desired fidelity, the pulse duration is increased accordingly.  Note that since the number 
of time steps, $N,$ is fixed, changing the pulse duration effectively changes $\Delta t$ and the nominal 
bandwidth must be scaled as $W/\Delta t$, so that the effective control bandwidth is kept constant. 

The results are shown in \figurename{~\ref{QSL-scaling}} with the 
mean of minimal times shown by the solid curve.  The dashed lines are  curve fits of the form  
$\frac{a}{W} +b$ showing approximate agreement with the mean. 
The filled region marks the standard deviation of minimal times across all samples. Some variance of minimal 
times is to be expected even when  restricting the target set to the perfect entanglers.  As shown in 
\cite{Watts-entropy}, the curvature of the SU(4) manifold  distorts the volume elements 
(Haar measure) throughout the Weyl chamber, effectively making some target unitaries {\it farther} 
from the initial identity matrix. From \figurename{~\ref{QSL-scaling}} it is clear that  the minimal time evolutions follow 
the time-bandwidth QSL scaling and demonstrate that Slepian based GRAFS not only recovers the bound of 
\cite{lloyd_qsl},  but also gives a constructive procedure for realizing the bound. 

\section{Conclusion}
The Slepian sequences, with their continuously varying bandwidth parameter, efficiently 
represent the space of band-limited controls. A gradient expression on the coefficients of a 
basis function expansion of the control was derived and shown to accurately and efficiently determine 
optimal controls.  Time optimal evolutions were numerically investigated and minimal pulse durations were
shown to scale with the inverse of the control bandwidth.  As quantum optimal control is increasingly 
applied to the experimental realm (cf. \cite{Heeres2017}),  GRAFS  may prove to be a useful 
technique for synthesizing band-limited time-optimal controls. Specific applications to trapped ion 
and superconducting qubit systems are left for future work. 

\section{Acknowledgements}This project was supported by the Intelligence Advanced
Research Projects Activity via Department of Interior National Business Center contract number 2012-12050800010. 
The author thanks Lorenza Viola and Michael J. Biercuk and Gregory Quiroz for helpful discussions. 

\bibliographystyle{apsrev4-1}
\bibliography{grafs_qsl}
\end{document}